\documentclass[fleqn,usenatbib]{mnras}

\usepackage{newtxtext, newtxmath, nicefrac, xcolor}
\usepackage[T1]{fontenc}

\DeclareRobustCommand{\VAN}[3]{#2}
\let\VANthebibliography\thebibliography
\def\thebibliography{\DeclareRobustCommand{\VAN}[3]{##3}\VANthebibliography}

\usepackage{graphicx}
\usepackage{amsmath}

\newcommand{\txd}{{\text{d}}}

\newcommand{\txp}{{\text{p}}}
\newcommand{\txt}{{\text{t}}}
\newcommand{\calE}{{\cal{E}}}
\newcommand{\calN}{{\cal{N}}}

\newcommand{\rT}{r_{\scriptscriptstyle{\text{T}}}}

\newcommand{\calET}{\calE_{\scriptscriptstyle{\text{T}}}}
\newcommand{\rhoT}{\rho_{\scriptscriptstyle{\text{T}}}}
\newcommand{\varrhoT}{\varrho_{\scriptscriptstyle{\text{T}}}}
\newcommand{\ra}{r_{\text{a}}}
\newcommand{\rb}{r_{\text{b}}}

\DeclareMathOperator{\arctanh}{arctanh}

\defcitealias{2022MNRAS.512.2266B}{Paper~I}

\title[Dynamical models with a finite extent~II]{Self-consistent dynamical models with a finite extent -- II. Radially truncated models}

\author[M. Baes]{Maarten Baes
\\
Sterrenkundig Observatorium, Universiteit Gent, Krijgslaan 281 S9, 9000 Gent, Belgium
}

\date{Accepted 9 January 2023. Received 22 December 2022; in original form 25 October 2022}

\pubyear{2022}

\begin{document}
\label{firstpage}
\pagerange{\pageref{firstpage}--\pageref{lastpage}}
\maketitle

\begin{abstract}
Galaxies, dark matter haloes, and star clusters have a finite extent, yet most simple dynamical models have an infinite extent. The default method to generate dynamical models with a finite extent is to apply an energy truncation to the distribution function, but this approach is not suited to construct models with a preset density profile and it imposes unphysical constraints on the orbit population. We investigate whether it is possible to construct simple dynamical models for spherical systems with a preset density profile with a finite extent, and ideally with a different range of orbital structures. We systematically investigate the consistency of radially truncated dynamical models, and demonstrate that no spherical models with a discontinuous density truncation can be supported by an ergodic orbital structure. On the other hand, we argue that many radially truncated models can be supported by a tangential Osipkov--Merritt orbital structure that becomes completely tangential at the truncation radius. We formulate a consistency hypothesis for radially truncated models with such an orbital structure, and test it using an analytical example and the numerical exploration of a large model parameter space using the {\tt{SpheCow}} code. We physically interpret our results in terms of the occupancy of bound orbits, and we discuss possible extensions of the tangential Osipkov--Merritt orbital structure that can support radially truncated models.
\end{abstract}


\section{Introduction}

In the study of galaxies, dark matter haloes, and star clusters, analytical spherical dynamical models are one of the most useful tools. They can serve as a first-order model to characterise these structures, they are useful as the starting point for full-scale numerical simulation, or they can act as a laboratory setting in which the effects of physical processes can be explored. The impressive range of applications of popular models such as the King models \citep{1966AJ.....71...64K}, the isochrone sphere \citep{1959AnAp...22..126H, 1960AnAp...23..474H}, the Hernquist model \citep{1990ApJ...356..359H}, the Plummer sphere \citep{1911MNRAS..71..460P}, the NFW model \citep{1997ApJ...490..493N, 2001MNRAS.321..155L}, or the S\'ersic model \citep{1968adga.book.....S, 1991A&A...249...99C} clearly illustrates their value.

In building spherical dynamical models, there are two main approaches \citep{2008gady.book.....B, 2021isd..book.....C}. The $f$-to-$\rho$ approach starts with an explicit expression for the phase-space distribution function $f(\calE,L)$, where $\calE$ and $L$ represent the binding energy and the angular momentum per unit mass, respectively. By selecting a distribution function that is positive for all values of $\calE$ and $L$, one is automatically ensured that the dynamical model is consistent, that is, physically viable. The disadvantage of this $f$-to-$\rho$ approach is that it does not lead to an explicit expression for any of the important dynamical properties such as the density or the velocity dispersion profiles. At best the density and potential can be obtained by numerically solving Poisson’s equation. As a result, this approach is not suited to construct models with a preset density profile, as one often wants to do.

The second approach, known as the $\rho$-to-$f$ approach, starts from an explicit expression for the density profile. The potential can be derived using Poisson's equation. With the assumption of an orbital structure, the distribution function and all other important dynamical properties can subsequently be calculated, at least in principle. The main challenge of this approach is that, except for a number specific choices for the orbital structure, the determination of the distribution function is far from trivial \citep{1986PhR...133..217D}. Moreover, not every density profile can be generated self-consistently by any orbital structure: only if the corresponding distribution function is positive over the entire phase space, the dynamical model is consistent. This is nearly impossible to know without actually calculating and investigating the distribution function. Examples demonstrate that the consistency is not always guaranteed, even for relatively simple density profiles and orbital structures \citep[e.g.,][]{2019A&A...626A.110B, 2021MNRAS.503.2955B}.

The vast majority of all analytical models presented in the literature have an infinite extent, meaning that they have a non-zero density at all radii. Dynamical systems such as galaxies or star clusters have a finite extent, so it is meaningful to investigate the possibility to build models with a finite extent. A popular way to do so is to apply a truncation in binding energy to the distribution function. By excluding the orbits with the lowest binding energies, one automatically excludes all particles or stars beyond a given truncation radius. The most famous example of this approach is the family of King models, which are energy-truncated versions of the isothermal sphere \citep{1963MNRAS.125..127M, 1966AJ.....71...64K, 2008gady.book.....B}. There are two major issues linked to this approach, however. The first one is that energy-truncated models, by definition, belong to the $f$-to-$\rho$ category, so the density profile cannot be set at the beginning, and all important dynamical properties can usually only be calculated numerically. The second disadvantage is that a truncation in binding energy imposes artificial and unphysical constraints on the system \citep{1988ApJ...325..566K}. Indeed, it prohibits a fraction of orbits in the model to be populated, even though these orbits are gravitationally bound to the system and remain within the allowed radial range. Especially nearly circular orbits near the truncation radius of the system are excluded when an energy truncation is applied. 

This raises the question whether it is possible to construct simple dynamical models for spherical systems with a preset density profile with a finite extent, and ideally with a different range of orbital structures. In \citet{2022MNRAS.512.2266B}, hereafter \citetalias{2022MNRAS.512.2266B}, we started our investigation into this question by looking in detail at the special case of the uniform density sphere, the simplest model with a radially truncated density profile. It was already well-known that the uniform density sphere cannot be supported by an ergodic orbital structure \citep{Zeldovich72, 1979PAZh....5...77O, 2008gady.book.....B}. We demonstrated that the uniform density sphere is also inconsistent with any constant anisotropy or radial Osipkov--Merritt orbital structure, but we constructed a family of self-consistent dynamical models for the uniform density sphere in which all possible orbits are populated. 

In this paper, the second in a series on self-consistent dynamical models with a finite extent, we systematically investigate the consistency of radially truncated dynamical models. More specifically, we start from an arbitrary spherical density profile $\rho_0(r)$ with an infinite extent, thus $\rho_0(r)>0$ for all $r$, and we create a new model by applying a radial truncation to this density profile, 
\begin{equation}
\rho(r) = \rho_0(r)\,\Theta(\rT-r), 
\label{truncdensity}
\end{equation}
where $\rT$ represents the truncation radius and $\Theta(x)$ the Heaviside step function. We stress that we specifically focus on models with a discontinuous truncation, so with $\rho(r)\ne0$ for $r\to\rT$. We want to investigate whether it is possible to build self-consistent dynamical models corresponding to this density profile, and if so, to find out which orbital structure would support it.

This paper is organised as follows. In Section~{\ref{general.sec}} we present a general consistency analysis for radially truncated models of the type~(\ref{truncdensity}). We start by considering ergodic models (Section~{\ref{iso.sec}}) and subsequently move on the type~II Osipkov--Merritt models (Section~{\ref{OM.sec}}). At the end of this section, we formule a consistency hypothesis for radially truncated models, which we test in Section~{\ref{test.sec}} using both an analytical case (Section~{\ref{TP.sec}}) and a general numerical approach (Section~{\ref{SpheCow.sec}}). In Section~{\ref{discussion.sec}} we discuss our findings, and we summarise in Section~{\ref{summary.sec}}.

\section{General consistency analysis}
\label{general.sec}

\subsection{Ergodic orbital structure}
\label{iso.sec}

To investigate whether the model defined by the density profile~(\ref{truncdensity}) can be supported self-consistently by an ergodic orbital structure, we need to calculate the unique ergodic distribution function $f(\calE)$ and check whether it is positive over the entire phase space. The ergodic distribution function can be calculated through the standard Eddington equation,
\begin{equation}
f(\calE) = \frac{1}{2\sqrt{2}\,\pi^2}\,\frac{\txd}{\txd\calE} \int_0^{\calE} \frac{\txd\tilde\rho}{\txd\Psi}\, \frac{\txd\Psi}{\sqrt{\calE-\Psi}},
\label{Eddington}
\end{equation}
where $\tilde\rho(\Psi)$ represents the augmented density, that is, the density written as a function of the potential \citep{1986PhR...133..217D}. Since the density is truncated at $r=\rT$, we find for the augmented density
\begin{equation}
\tilde\rho(\Psi) = \tilde\rho_0(\Psi)\,\Theta(\Psi-\calET), 
\label{tilderho}
\end{equation}
with the truncation energy $\calET$ defined as $\calET = \Psi(\rT)$. The formal derivative is
\begin{equation}
\frac{\txd\tilde\rho}{\txd\Psi}(\Psi) = \tilde\rho_0'(\Psi)\,\Theta(\Psi-\calET) + \rhoT\,\delta(\Psi-\calET),
\label{formder}
\end{equation}
where we have introduced the notation
\begin{equation}
\rhoT = \rho_0(\rT) = \tilde\rho_0(\calET) > 0.
\end{equation}
Inserting expression (\ref{formder}) in Eddington's equation and applying partial integration, we obtain
\begin{multline}
f(\calE) 
= 
\frac{\rhoT}{2\sqrt2\, \pi^2}
\frac{\delta(\calE-\calET)}{\sqrt{\calE-\calET}}
- \frac{\Theta(\calE-\calET)}{2\sqrt2\, \pi^2}\left[ 
\frac{\rhoT}{2\,(\calE-\calET)^{3/2}} 
\right.
\\
\left.
+ \frac{\rT\,\rho'_0(\rT)}{\calET \sqrt{\calE-\calET}}
- \int_{\calET}^\calE \frac{\tilde\rho_0''(\Psi)\,\txd\Psi}{\sqrt{\calE-\Psi}}
\right],
\label{dftrunc}
\end{multline}
where we have used the fact that
\begin{equation}
\tilde\rho_0'(\calET) 
=
\frac{\rT\,\rho_0'(\rT)}{\calET}.
\end{equation}
Expression~(\ref{dftrunc}) shows several interesting characteristics. Firstly, the distribution function is identically zero for $\calE<\calET$, which means that orbits with binding energy below the truncation energy $\calET$ are not populated. A radial truncation of the density profile thus automatically generates a truncation in binding energy for the ergodic distribution function. Secondly, the first term in (\ref{dftrunc}) contains a Dirac delta function at the truncation energy. In principle, a Dirac delta function term in the distribution function is not necessarily an issue, but the fascinating aspect here is that the weight is infinite. Finally, the term between the square brackets in Eq.~(\ref{dftrunc}) is always negative at binding energies just beyond the truncation energy. Indeed, the first term will dominate the distribution function for $\calE\gtrsim\calET$, and since $\rhoT>0$, we find that the distribution function is negative. We thus find that all spherical models with a discontinuous density truncation have inconsistent ergodic distribution functions, that is, they cannot be supported by an ergodic orbital structure. 

\subsection{Tangential Osipkov--Merritt orbital structure}
\label{OM.sec}

The fact that truncated models cannot be supported by an ergodic orbital structure does not imply that it is impossible to generate self-consistent dynamical models for them. Some orbital configurations are less demanding than others. In particular, tangentially anisotropic dynamical models are generally less demanding than radially anisotropic ones. If we want to search for positive distribution functions, and thus physically viable dynamical models for truncated density profiles, it seems wise we have to focus on orbital structures with tangential anisotropy.

One interesting option is the class of tangential Osipkov--Merritt models. These models, denoted as type II models by \citet{1985AJ.....90.1027M}, are characterised by an ellipsoidal velocity distribution and an anisotropy profile that gradually changes from isotropic in the centre to completely tangential at the radius $\ra$, the anisotropy radius. More specifically, 
\begin{equation}
\beta(r) = -\frac{r^2}{\ra^2-r^2}.
\label{TOMbetagen}
\end{equation}
These type II or tangential Osipkov--Merritt (hereafter TOM) models are only meaningful for models with a finite extent, with $\ra\geqslant \rT$. In \citetalias{2022MNRAS.512.2266B} we showed that the uniform density sphere cannot be supported by an ergodic orbital structure, but that it can be supported by a TOM orbital structure with $\ra = \rT$.

To investigate whether TOM models are a valid option for general truncated density models, we need to go through a similar analysis as for the ergodic case: we need to calculate the distribution function and check the positivity over the entire phase space. The distribution function of TOM models depends on binding energy $\calE$ and angular momentum $L$ only through the combination
\begin{equation}
Q = \calE + \frac{L^2}{2\ra^2}.
\label{Q-}
\end{equation}
Given a density profile and anisotropy radius, it can be obtained using an inversion relation similar to Eq.~(\ref{Eddington}),
\begin{equation}
f(\calE,L) \equiv f(Q) = \frac{1}{2\sqrt2\,\pi^2}\,\frac{\txd}{\txd Q} \int_0^Q \frac{\txd\varrho}{\txd\Psi}\,
\frac{\txd\Psi}{\sqrt{Q-\Psi}},
\label{om}
\end{equation}
with
\begin{equation}
\varrho(\Psi) = \left. \left(1-\frac{r^2}{\ra^2}\right) \rho(r)\,\right|_{r=r(\Psi)}.
\label{hom}
\end{equation}
With a density profile as in Eq.~(\ref{truncdensity}), the function $\varrho(\Psi)$ reads
\begin{equation}
\varrho(\Psi) = \varrho_0(\Psi)\,\Theta(\Psi-\calET).
\label{tildevarrho}
\end{equation}
Comparing the expressions~(\ref{om}) and (\ref{tildevarrho}) to the expressions~(\ref{Eddington}) and (\ref{tilderho}), we see that we have formally identical equations as in the ergodic case if we make the substitutions $\tilde\rho\mapsto \varrho$ and $\calE\mapsto Q$. As a result, we can immediately write for the distribution function,
\begin{multline}
f(Q) 
= 
\frac{\rhoT}{2\sqrt2\, \pi^2}
\frac{\delta(Q-\calET)}{\sqrt{Q-\calET}}
- \frac{\Theta(Q-\calET)}{2\sqrt2\, \pi^2}\left[ 
\frac{\varrhoT}{2\,(Q-\calET)^{3/2}} 
\right.
\\
\left.
+ \frac{\varrho'_0(\calET)}{\sqrt{Q-\calET}}
- \int_{\calET}^Q \frac{\varrho_0''(\Psi)\,\txd\Psi}{\sqrt{Q-\Psi}}
\right].
\label{fQgen1}
\end{multline}
with 
\begin{equation}
\varrhoT = \varrho(\calET) = \left(1-\frac{\rT^2}{\ra^2}\right) \rhoT.
\end{equation}
In the general case with $\ra>\rT$, we have $\varrhoT>0$, and we can, based on the discussion in the previous subsection, conclude that the distribution function (\ref{fQgen1}) is not positive over the entire phase space. The TOM dynamical models with $\ra>\rT$ are thus physically inconsistent.

The situation changes for the special case $\ra=\rT$, however. In this case the anisotropy changes systematically from isotropy in the centre to complete tangential anisotropy at the truncation radius,
\begin{equation}
\beta(r) = -\frac{r^2}{\rT^2-r^2}.
\label{TOMbeta}
\end{equation}
Eq.~(\ref{tildevarrho}) shows that $\varrhoT=0$ in this case, and as a result the distribution function simplifies to
\begin{equation}
f(Q) 
= 
\frac{\Theta(Q-\calET)}{2\sqrt2\, \pi^2}
\left[
\frac{2 \rhoT}{\calET \sqrt{Q-\calET}}
+
\int_{\calET}^Q \frac{\varrho_0''(\Psi)\,\txd\Psi}{\sqrt{Q-\Psi}}
\right].
\label{fQ}
\end{equation}
Compared to the general case~(\ref{fQgen1}), both the term with the Dirac delta function and the term that diverges negatively for $Q\gtrsim\calET$ have disappeared. 

Unfortunately, it is not immediately possible to make firm general statements about the positivity of this expression over the entire phase space: whether or not this expression is positive for all values of $Q$ depends on the specific shape of the density profile and on the value of the truncation radius. On the other hand, it is clear that the expression~(\ref{fQ}) has the potential to be a positive and hence physically viable distribution function for large classes of truncated density models. For the smallest values of $Q$ for which the distribution is non-zero, $Q\gtrsim\calET$, first term between the square brackets will dominates the expression, and this term is always positive. For the largest values of $Q$, corresponding to the smallest radii, the TOM models are isotropic and we expect a similar behaviour as the ergodic distribution function of the non-truncated model.

Based on these considerations, we formulate the following consistency hypothesis: 
\newline\newline
{\it{If a spherical density model can be supported self-consistently by an ergodic orbital structure, then the model that is radially truncated at $r=\rT$ can be supported self-consistently by the TOM orbital structure with $\ra$ = $\rT$. }}

\section{Testing the consistency hypothesis}
\label{test.sec}

In this section we test the consistency hypothesis we formulated at the end of the previous section. We first present a family of truncated density models for which we can analytically calculate the distribution function under the assumption of a TOM orbital structure. This allows an explicit investigation of the consistency of these dynamical models. Subsequently we will explore a larger suite of models using numerical means.

\subsection{A completely analytical model}
\label{TP.sec}

As our first test case we consider a truncated version of the \citet{1911MNRAS..71..460P} sphere, one of the most popular models in stellar dynamics studies. This model is characterised by the density profile
\begin{equation}
\rho_0(r) = \frac{3}{4\pi}\,\frac{M}{b^3} \left(1+\frac{r^2}{b^2}\right)^{-5/2}, 
\end{equation}
with $M$ the total mass, and $b$ a scale length. In the remainder of this section we use dimensionless units in which $G=M=b=1$, which simplifies the density profile to
\begin{equation}
\rho_0(r) = \frac{3}{4\pi}\left(1+r^2\right)^{-5/2}.
\label{Plummerrho0}
\end{equation}
The Plummer model is famous for having a simple power-law ergodic distribution function \citep{2008gady.book.....B}, and for the fact that the distribution function can also be calculated analytically for various other orbital configurations \citep{1979PAZh....5...77O, 1985AJ.....90.1027M, 1986PhR...133..217D, 1987MNRAS.224...13D, 1991MNRAS.253..414C, 2007A&A...471..419B}. 

When we apply a radial truncation to the density profile~(\ref{Plummerrho0}) we obtain
\begin{equation}
\rho(r) = \frac{3}{4\pi} \left(1+r^2\right)^{-5/2} \Theta(\rT-r).
\label{poii}
\end{equation} 
Introducing the notation
\begin{equation}
s = \frac{1}{\sqrt{1+\rT^2}},
\label{defs}
\end{equation}
the potential of this truncated Plummer model can, for $r\leqslant\rT$, be written as
\begin{equation}
\label{PluPsi}
\Psi(r)
=
\frac{1}{\sqrt{1+r^2}} - s^3.
\end{equation}
Combining the expressions~(\ref{hom}), (\ref{Plummerrho0}) and  (\ref{PluPsi}), we can immediately derive a compact expression for the function $\varrho_0(\Psi)$,
\begin{equation}
\varrho_0(\Psi) 
= 
\frac{3}{4\pi\,(1-s^2)} \left[ \bigl(\Psi+s^3\bigr)^5 - s^2\,\bigl(\Psi+s^3\bigr)^3 \right].
\end{equation}
Inserting this expression into the general expression~(\ref{fQ}) we find, after some algebra, an explicit expression for the TOM distribution function,
\begin{equation}
f(Q) 
= 
\frac{3\sqrt2}{56\pi^3}\,\frac{\Theta(Q-\calET)}{\,(1-s^2)}
\left[ 
\frac{7s^4}{\sqrt{Q-\calET}}
+V(Q)\,\sqrt{Q-\calET}
\right],
\label{TP-dfOM}
\end{equation}
with $V(Q)$ a cubic polynomial,
\begin{multline}
V(Q) = 2 \left(4Q+3s+4s^3\right)
\\
\times
\left[8Q^2 - 2s\left(1-8s^2\right)Q + s^2\left(1-2s^2+8s^4\right) \right]
\end{multline}
and with
\begin{gather}
\calET = s\left(1-s^2\right).
\label{calETs}
\end{gather}

\begin{figure}
\centering
\includegraphics[width=0.95\columnwidth]{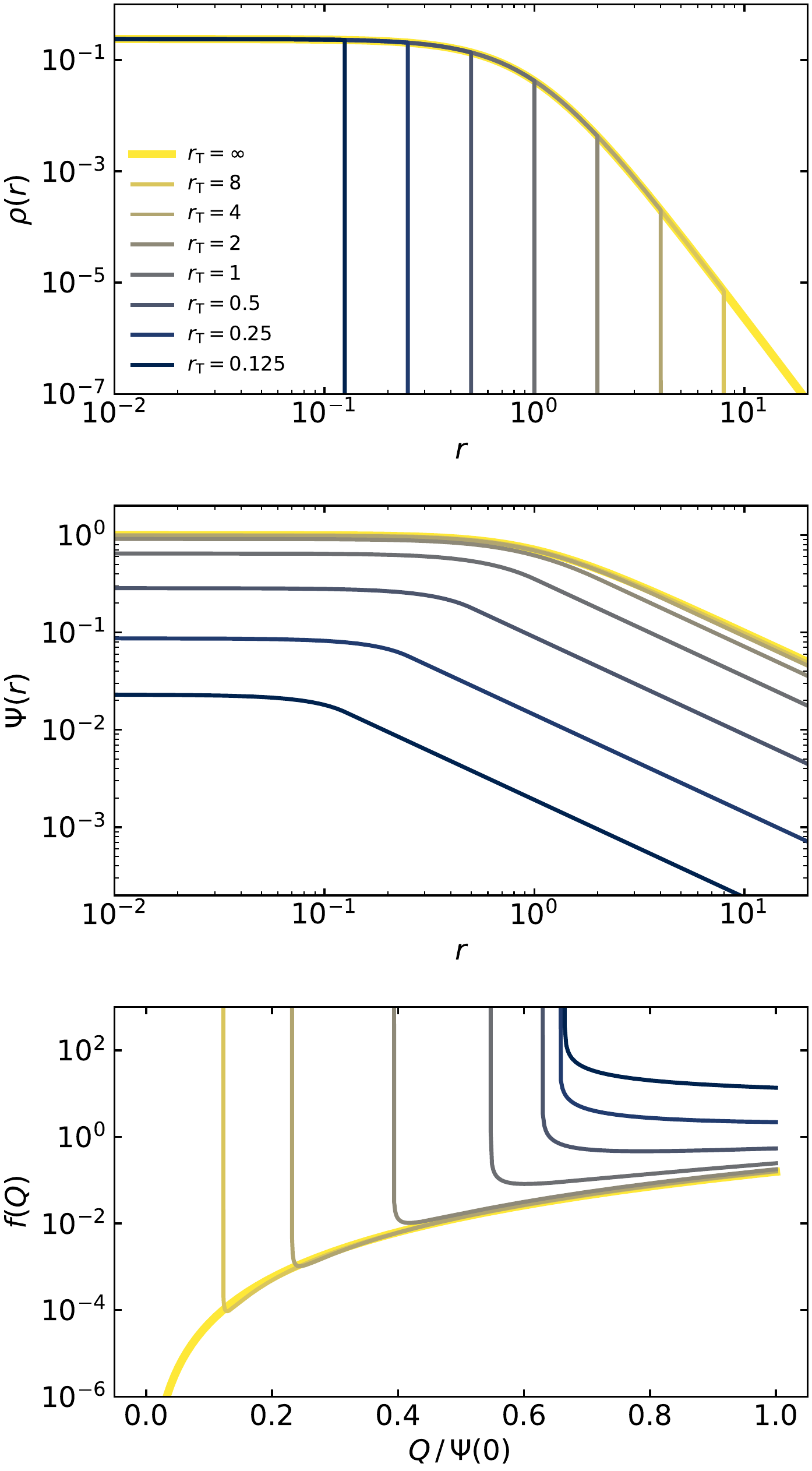}\hspace*{2em}
\caption{Some properties of the family of truncated Plummer models for different values of the truncation radius $\rT$. The different panels show the density (upper), potential (middle), and distribution function (bottom). We have chosen dimensionless units with $G=M=b=1$.}
\label{TruncatedPlummer.fig}
\end{figure}

In Fig.~{\ref{TruncatedPlummer.fig}} we show a number of characteristics of the family of truncated Plummer models for different values of the truncation radius. The upper panel shows the density, which is identical for all models up to the truncation radius, where it abruptly truncated. As a result of this truncation, the total mass is reduced and the gravitational potential (middle panel) is a decreasing function of $\rT$ at every radius. The lower panel shows the distribution function~(\ref{TP-dfOM}) as a function of $Q$, normalised to the central value of the potential. For each value of $\rT$, the distribution function is zero for all $Q<\calET$, it diverges towards infinity when $Q$ approaches $\calET$ from the high binding-energy side, and is positive for all $\calET<Q<\Psi(0)$. The most important characteristic is that the distribution function is non-negative for all $Q$, that is, over the entire phase space. This means that the truncated Plummer model can be supported by a TOM orbital structure. 

We note that the behaviour described above applies to any value of the truncation radius: all truncated Plummer models can be can be supported by a TOM orbital structure. Looking at the systematic behaviour as a function of truncation radius, we note that all quantities shown tend to converge to the yellow curves when $\rT$ increases. These yellow curves correspond to the limiting case $\rT\to\infty$, or no truncation at all. When we set $\rT=\ra\to\infty$ in Eqs.~(\ref{Q-}), (\ref{defs}), and (\ref{calETs}), we have $Q=\calE$, $s=0$ and $\calET=0$, and the distribution function~(\ref{TP-dfOM}) reduces to
\begin{equation}
f(\calE) = \frac{24\sqrt2}{7\pi^3}\,\calE^{7/2}.
\label{dfP}
\end{equation}
This distribution function is nothing but the ergodic distribution function of the non-truncated Plummer model \citep{1987MNRAS.224...13D, 2008gady.book.....B}. 

\subsection{Numerical investigation using {\tt{SpheCow}}}
\label{SpheCow.sec}

The Plummer model is quite unique in that the TOM distribution function of the truncated version can be calculated analytically, and the positivity can therefore easily be tested. For the vast majority of spherical density profiles this is not the case. In order to further test our consistency hypothesis, we resort to a numerical investigation. 

\subsubsection{Implementation in {\tt{SpheCow}}}

{\tt{SpheCow}} \citep{2021A&A...652A..36B} is a software tool designed to numerically explore the dynamical structure of any spherical model defined by an analytical density profile or surface density profile.  The code contains implementations for many commonly used models, including the Plummer, Hernquist, Jaffe, NFW, S\'ersic, Nuker, Einasto, and Zhao models, and is set up in such a way that new models can easily be added. For each model, the user can numerically explore the most important dynamical quantities, such as velocity dispersion profiles, the distribution function and the differential energy distribution, under the assumption of either an ergodic or a radially anisotropic Osipkov--Merritt orbital structure. Recent applications of {\tt{SpheCow}} include detailed analyses of the families of S\'ersic, Nuker, and Einasto models \citep{2019A&A...626A.110B, 2019A&A...630A.113B, 2020A&A...634A.109B, 2022A&A...667A..47B}, a study of the physical consistency of the double and broken power-law models \citep{2021MNRAS.503.2955B}, and an investigation of the differential energy distribution and the total integrated binding energy of dynamical models \citep{2021A&A...653A.140B}.

For the purpose of testing our consistency hypothesis for truncated models, two extensions to {\tt{SpheCow}} were required. The first extension was an implementation of the truncated version of large suite of density models. Secondly, as the code only considered ergodic and radially anisotropic Osipkov--Merritt models, we needed to expand its applicability to TOM orbital structures.

The first task was relatively straightforward. Rather than implementing a truncated version for every single existing, and future, model in {\tt{SpheCow}}, we used a generic approach based on the decorator design pattern. In object-oriented programming, a decorator is a design pattern that dynamically attaches additional responsibilities to an object \citep{Gamma1994, Freeman2004}. It provides a flexible and powerful alternative to subclassing for extending functionality. In the {\tt{SpheCow}} context, we defined a new {\tt{TruncatedModel}} class that takes another model as input and truncates its density profile at a user-specified truncation radius. The advantage of the decorator approach is clear: we only had to implement the {\tt{TruncatedModel}} decorator class once, and it can be applied to any possible model. It can also be applied to models without an analytical density profile, that is, models with an analytical surface density profile as starting point, such as the S\'ersic or Nuker models.

For the numerical calculation of the dynamical structure of an arbitrary spherical model under the assumption of a TOM orbital distribution, we needed to implement the relevant formulae, and convert them to integrations with respect to radius. Most of these formulae are modest adaptations of the formulae corresponding to a radial Osipkov--Merritt orbital distribution, as presented in Section~2.1.3 of \citet{2021A&A...652A..36B}, based on expressions derived by various authors \citep[e.g.,][]{1979PAZh....5...77O, 1985AJ.....90.1027M, 1991MNRAS.253..414C, 2002A&A...393..485B, 2005MNRAS.362...95M, 2014MNRAS.442.3284A, 2021isd..book.....C}. In Appendix~{\ref{TOM.sec}} we give explicit expressions for the most important dynamical quantities as implemented in {\tt{SpheCow}}.

\begin{figure}
\centering
\includegraphics[width=0.95\columnwidth]{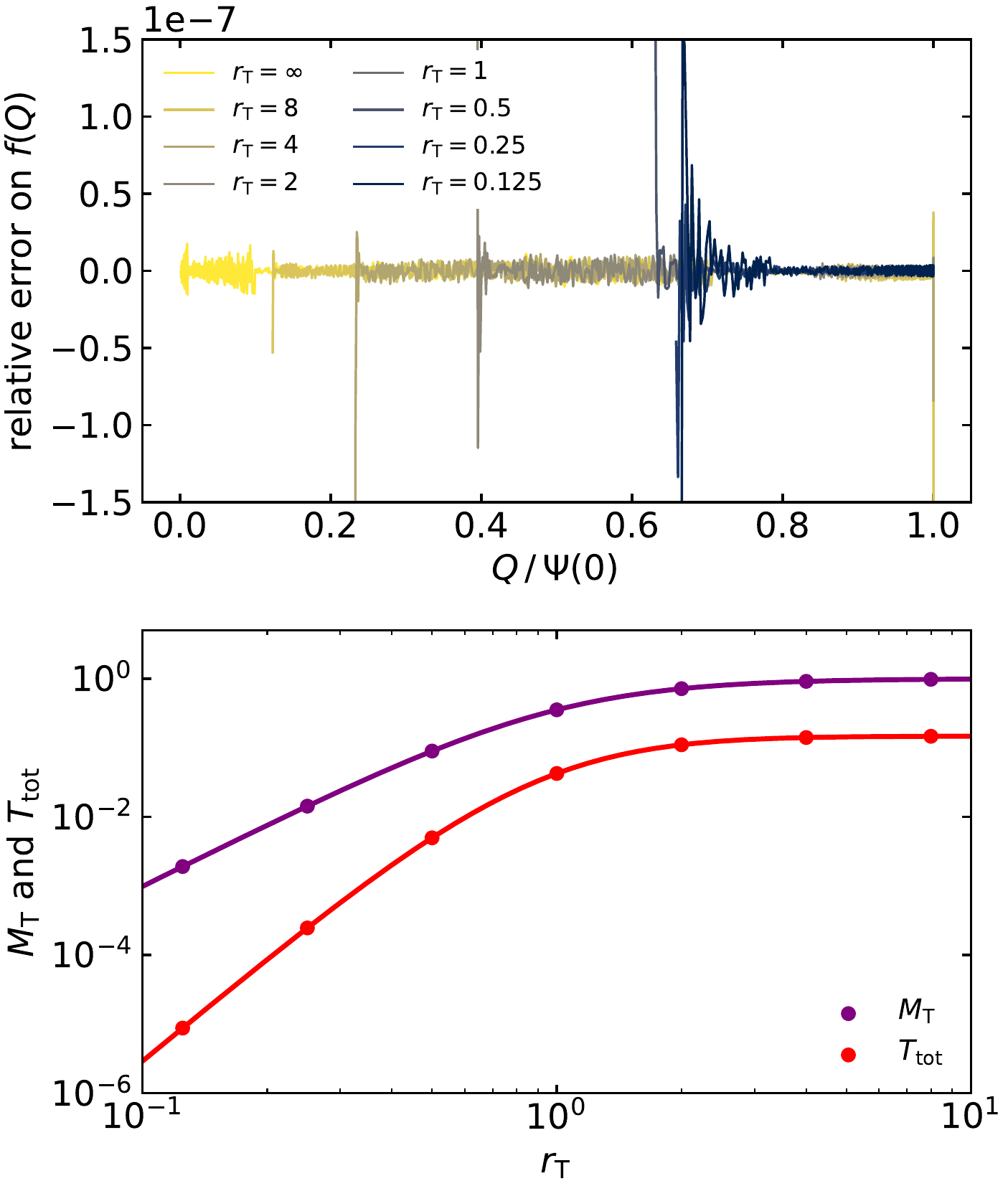}\hspace*{2em}
\caption{Convergence tests for the implementation of the truncation decorator and the TOM orbital structure within {\tt{SpheCow}}. The top panel shows the relative error between the analytical value and the {\tt{SpheCow}} value for the TOM distribution function of the truncated Plummer model, for the same truncation radii as shown in Fig.~{\ref{TruncatedPlummer.fig}}. The bottom panel shows the total mass and the total kinetic energy of the truncated Plummer model as a function of the truncation radius. Dots show the results of the {\tt{SpheCow}} calculation (see text for details), solid lines are the analytical results. We have chosen dimensionless units with $G=M=b=1$.}
\label{TestSpheCow.fig}
\end{figure}

To check the accuracy of the new implementations, we have compared the results of the {\tt{SpheCow}} calculations for the truncated Plummer model to the analytical results shown in Fig.~{\ref{TruncatedPlummer.fig}}. The top panel shows the relative error on the distribution function. For 128 Gauss-Legendre integration points, as suggested by \citet{2021A&A...652A..36B} to be a good compromise between accuracy and speed, we reproduced the analytical distribution function with a typical root mean square relative error of the order of $10^{-8}$. The bottom panel shows a comparison involving more complex calculations. The purple data in Fig.~{\ref{TestSpheCow.fig}} shows the remaining mass of the truncated Plummer model as a function of the truncation radius. The solid line shows the analytical result, whereas the dots show the {\tt{SpheCow}} calculation, obtained by integrating the pseudo-differential energy distribution $N(Q)$ over all values of $Q$. Even the calculation of this quantity, which contains various nested levels of numerical quadrature, is accurate to at least 8 significant digits. The red data show the calculation of the total kinetic energy as a function of $\rT$, and the solid line represents the analytical result. The dots represent the {\tt{SpheCow}} results, obtained by integrating the velocity dispersion profiles, which by themselves involve two levels of quadrature. Also here, we easily recover the analytical results to at least 8 significant digits.

\subsubsection{Exploration of the model space}

\begin{figure*}
\centering
\includegraphics[width=0.96\textwidth]{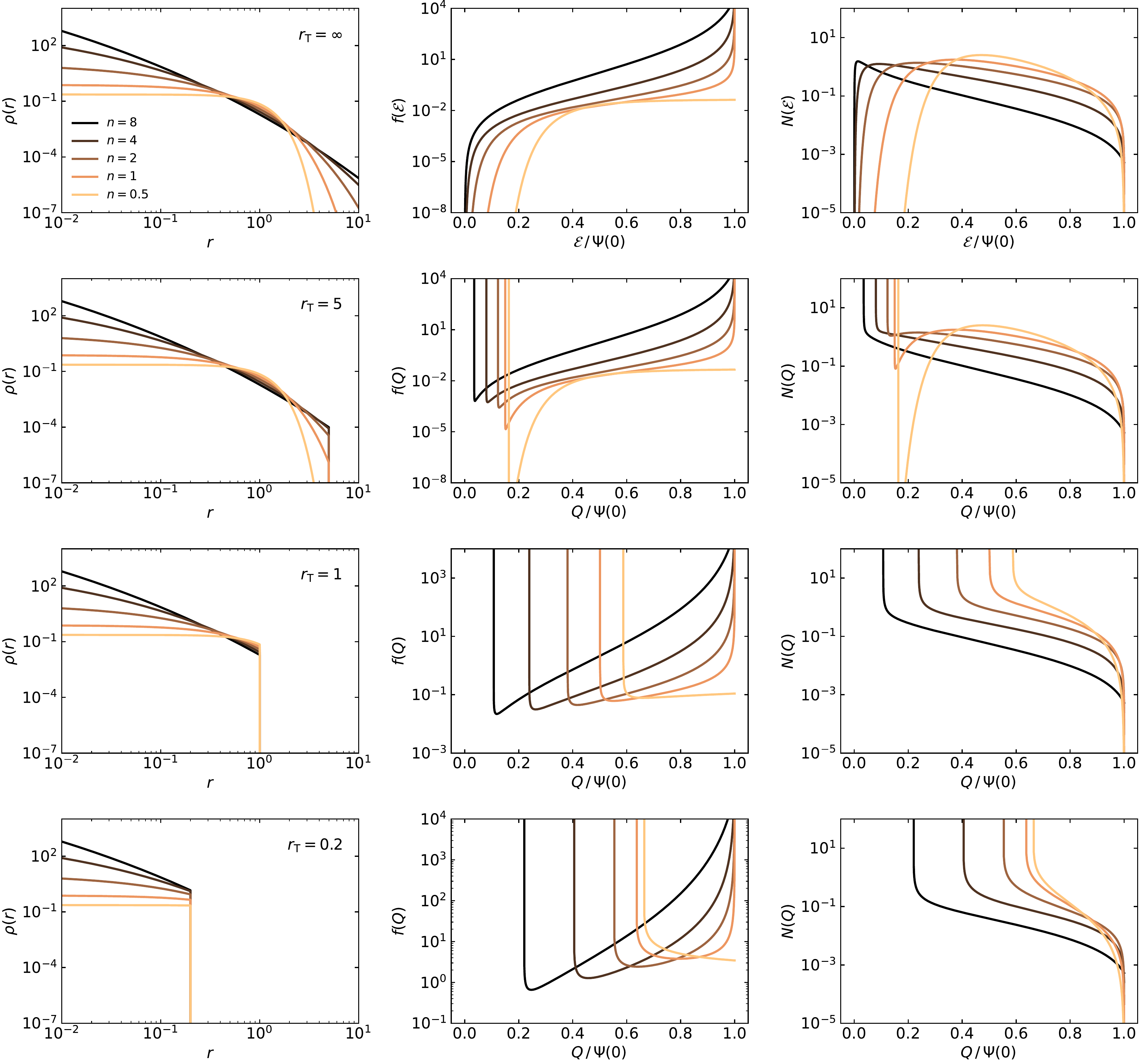}\hspace*{4em}
\caption{Density (left column), distribution function (middle column), and pseudo-differential energy distribution (right column) for a set of truncated Einasto models, assuming a TOM orbital structure. The different rows correspond to decreasing values of $\rT$, with the top row corresponding to the non-truncated models ($\rT\to\infty$) and the bottom row to model truncated at small radii ($\rT=0.2$). Within each panel, different colours correspond to Einasto models with different values for the Einasto index, $n$, as indicated in the upper left panel. We have adopted dimensionless units with $G=M=r_{\text{h}}=1$.}
\label{TruncatedEinasto.fig}
\end{figure*}

The extended {\tt{SpheCow}} code allows a numerical investigations of the consistency hypothesis formulated at the end of Section~{\ref{OM.sec}}.  Our approach was relatively straightforward: for each model implemented in {\tt{SpheCow}} for which we know that the ergodic distribution function is positive for all binding energies, we constructed different truncated versions by selecting different truncation radii. Subsequently we calculated the corresponding TOM distribution function and we checked its positivity over all $\calET<Q<\Psi_0$. At the same time, we calculated some other dynamical quantities, such as the velocity dispersion and the pseudo-differential energy distribution, which were used to test the validity of the calculations using the consistency checks described in the previous subsection.

Fig.~{\ref{TruncatedEinasto.fig}} presents an example of this approach for the family of Einasto models. The structure and dynamics of this family of models has been investigated by various authors \citep[e.g.,][]{2005MNRAS.362...95M, 2005MNRAS.358.1325C, 2010MNRAS.405..340D, 2012A&A...540A..70R, 2012A&A...546A..32R}. Using the {\tt{SpheCow}} code, \citet{2022A&A...667A..47B} showed that the Einasto model has a positive ergodic distribution function for all models with Einasto index $n\geqslant\tfrac12$. Fig.~{\ref{TruncatedEinasto.fig}} show the density, distribution function, and pseudo-differential energy distribution for Einasto models with different Einasto indices (different colours) and different truncation radii: the top row corresponds to the non-truncated Einasto model ($\rT\to\infty$), and the subsequent row to gradually decreasing values ($\rT=5$, 1 and 0.2). These three values have been chosen as representative for the three regimes: they correspond to a truncation beyond, at, and before the half-mass radius, respectively.

In all cases, we see that the behaviour of the dynamical properties at small radii ($r\ll\rT$) mimics the behaviour of the corresponding non-truncated ergodic Einasto models. In particular, we find that the distribution function for the limiting $n=\tfrac12$ model, corresponding to a truncated Gaussian density profile, converges to a finite, non-zero value for $Q\to\Psi(0)$ for all values of $\rT$, whereas all models with larger Einasto index have a distribution function that diverges as $Q$ approaches $\Psi(0)$. This similarity is logical as all models share the same density profile for $r<\rT$, and the TOM models are also isotropic at small radii ($r\ll\rT$). The similarity at small radii between the TOM truncated models and the ergodic non-truncated models decreases for decreasing $\rT$, as can be expected. 

At larger radii, on the other hand, we find a radically different behaviour between the truncated and non-truncated models. For the standard Einasto models, the ergodic distribution function is a positive and monotonically increasing function of binding energy over the entire range $0<\calE<\Psi(0)$, and for $\calE\to0$ we find that the distribution function smoothly converges to zero. For the truncated models, the distribution function is identically zero for all $0<Q<\calET$, and it diverges as $(Q-\calET)^{-1/2}$ for $Q\gtrsim\calET$, as expression~(\ref{fQ}) indicates. A similar radical change in behaviour is seen at large radii when comparing the differential energy distribution for the isotropic models and the pseudo-differential energy distribution of the corresponding truncated models.

The main result of this numerical investigation is that all truncated Einasto models with $n\geqslant\tfrac12$ have a distribution function that is positive over the entire range of $Q$, for each value of the truncation radius. These truncated models can thus be supported by a TOM orbital structure with $\ra=\rT$. 

We have repeated this exercise for different models with a positive ergodic distribution function, including the families of Dehnen or $\gamma$--models \citep{1993MNRAS.265..250D, 1994AJ....107..634T},  S\'ersic models \citep{1991A&A...249...99C, 2019A&A...626A.110B}, and generalised NFW models \citep{2000ApJ...529L..69J, 2006AJ....132.2685M}. We consistently find the same result: for the range of parameters for which these models have a positive ergodic distribution function, all  the truncated counterparts can be supported by a TOM orbital structure.

\section{Discussion}
\label{discussion.sec}

\subsection{Physical interpretation: orbit occupancy}
\label{interpret.sec}

The main result of this paper is that radially truncated spherical models with a density discontinuity can never be supported self-consistently by an isotropic orbital structure, whereas they can often be supported by a TOM orbital structure with $\ra=\rT$. In Section~{\ref{general.sec}} we have provided a mathematical proof, but we have not yet provided a physical interpretation.

To understand these results physically, it is useful to view a dynamical model as a combination of orbits, essentially as in Schwarzschild's orbit superposition method \citep[e.g.,][]{1984ApJ...286...27R, 2021MNRAS.500.1437N}. Each orbit in a spherical potential is a plane rosetta orbit, and we group all the orbits with the same pericentre and apocentre into a single building block (which we still call an orbit). The construction of a dynamical model comes down to determining the weight of each orbit. When building up a spherical dynamical model in this way, it is most convenient to start at the outermost radius and to gradually add orbits to the mixture with continuously decreasing apocentres. For each apocentre $r_+$, we have a range of orbits to choose from, corresponding to different pericentres $r_-$, and we have to determine the weight of each of them by ensuring that both the resulting density and velocity distribution at $r=r_+$ are in agreement with the predetermined density and orbital structure. 

When viewing a dynamical model in this way, it might seem strange that some distribution functions are negative. The reason is that each orbit does not only contribute to the density at its apocentre, but it also pollutes the density at smaller radii, and these contributions need to be taken into account when calculating the density and velocity distribution. Adding orbits with a given apocentre to reproduce the constraints at that radius can sometimes create an excess at smaller radii that can only be lifted by adding orbits with a negative weight, which obviously does not lead to a physically valid dynamical model. 

Let us now concentrate on models with a discontinuous radial density truncation, and specifically look at the ergodic orbital structure. If we want to build up an ergodic distribution function, we can only add families of orbits to the mixture, all with the same binding energy. If we start building up these models outside-in, we first need to add the family of orbits that reaches $\rT$ at the apocentre of the radial member of this family, which corresponds to the binding energy $\calE=\calET$. This ergodic family of orbits has a distribution function $f(\calE) = c\,\delta(\calE-\calET)$, with $c$ a constant that sets the weight of this family. The density profile corresponding to this distribution function is
\begin{multline}
\rho(r) = 4\sqrt2\, \pi \int_0^{\Psi(r)} f(\calE)\, \sqrt{\Psi(r)-\calE}\,\txd\calE 
\\
= 4\sqrt2\,\pi\, c\, \sqrt{\Psi(r)-\calET}\,\Theta(\rT-r).
\end{multline}
None of the orbits extends beyond $\rT$, as required. On the other hand, we note that this density profile smoothly converges to zero at $r=\rT$. Since all other families of orbits with $\calE_i<\calET$ do not contribute to the density at the truncation radius, the only way to realise a sharp density truncation at $r=\rT$ is to give the family above an infinite weight, that is, by including an infinite number of stars in the mixture with binding energy $\calET$. This will obviously create an infinite density at all radii $r<\rT$ as well, which can only be alleviated by adding orbits with negative weights, such that the final sum is finite. This combination of positive and negative contributions can be seen in Eq.~(\ref{dftrunc}): the first term contains the infinite positive contribution at the discrete binding energy $\calE=\calET$, whereas the second term accounts for the negative contribution (and potentially some positive contributions as well). These negative weights imply that the dynamical model is non-physical.

The same situation arises for many other orbital structures: not only ergodic orbital structures will fail to support radially truncated density models, but actually most orbital structures. In Section~{\ref{OM.sec}} we demonstrated that they cannot be supported by TOM orbital structures with $\ra>\rT$, and the same argumentation goes for, for example, models with constant anisotropy or a radial Osipkov--Merritt orbital structure. For the specific case of the uniform density sphere, this was demonstrated explicitly in \citetalias{2022MNRAS.512.2266B}. 

For TOM models with $\ra=\rT$, we can start populating the model with a mix of orbits with $r_+=\rT$ without immediate problems, and we can gradually work our way inwards, while attempting to reproduce both the density and the velocity distribution at each radius. The examples shown throughout this paper demonstrate that many truncated spherical density profiles can be generated self-consistently by a TOM orbital structure with $\ra=\rT$.

\subsection{The consistency hypothesis}

At the end of Section~{\ref{OM.sec}} we formulated a consistency hypothesis that states that any truncated density model for which the non-truncated counterpart can be supported by an ergodic orbital structure, can be supported by the TOM orbital structure with $\ra=\rT$. A number of reflections on this consistency hypothesis are appropriate.

First of all, we emphasise that this hypothesis remains an hypothesis. We tested its validity by numerically calculating the TOM distribution functions for a large set of such models, and consistently obtained the same result. This strengthens our conviction that it is generally valid, but it is not conclusive evidence. Attempts to formally proof the positivity of the TOM distribution function, or more specifically the second term between the square brackets in Eq.~(\ref{fQ}), given the positivity of the ergodic distribution function of the non-truncated model, did not lead to useful results.

Secondly, it is useful to indicate that our consistency hypothesis is a sufficient but not a necessary criterion. It is easy to generate truncated density models that can be supported by a TOM orbital structure while the non-truncated version cannot be supported by an ergodic distribution function. A simple example is the uniform density sphere, which can be supported by a TOM orbital structure \citepalias{2022MNRAS.512.2266B}. The uniform density sphere is, however, also the truncated version of any broken power-law model with $\gamma=0$, and none of these models can be supported by an ergodic orbital structure \citep{2021MNRAS.503.2955B}. Interestingly, the TOM orbital structure can potentially even support truncated density models with a central hole. It is well-known that the density needs to be a monotonically decreasing function of radius for a model to be supported by an ergodic distribution function, which can be considered as a special case of the global density slope--anisotropy inequality \citep[GDSAI;][]{2009MNRAS.393..179C, 2010MNRAS.408.1070C},
\begin{equation}
\gamma(r) \equiv -\frac{\txd\ln\rho}{\txd\ln r}(r) \geqslant 2\beta(r).
\end{equation}
The GDSAI is shown to be satisfied for all dynamical models with a separable augmented density for which the central anisotropy $\beta_0\leqslant\tfrac12$ \citep{2011ApJ...726...80V, 2011MNRAS.413.2554A, 2011ApJ...736..151A}. For the models with a TOM orbital structure, which belong to this class of models, the GDSAI becomes
\begin{equation}
\gamma(r) \geqslant -\frac{2r^2}{\rT^2-r^2}.
\end{equation}
All models with a monotonically decreasing density profile automatically satisfy this criterion, since the right-hand side of this inequality is always negative. Interestingly, the GDSAI leaves the door open for models in which $\gamma(r)$ is not necessarily positive over the entire radial range, including models with a central density hole. 

A third and final reflection on the consistency hypothesis is that not all truncated density models can be supported self-consistently by a TOM orbital structure. It is easy to generate counter-examples. For example, \citet{2022A&A...667A..47B} showed that Einasto models with Einasto index $n<\tfrac12$ cannot be supported by an isotropic orbital structure. These models are created by a very flat density profile at small radii, followed by a very sharp (but not infinitely sharp) break in the density profile. If this model is truncated at a sufficiently small radius, it essentially reduces to the uniform density sphere, which is compatible with a TOM orbital structure. Truncating it at a radius beyond the break radius results in a model that cannot be supported self-consistently by a TOM orbital structure.

\subsection{Alternative orbital structures}

The TOM orbital structure with $\ra = \rT$ is a viable option that can support many, but not all, truncated density profiles. On top of that, the TOM orbital structure has the very useful characteristic that the determination of the distribution function is relatively simple: it can be derived from the density using an Eddington-like inversion formula involving just a single integral. It is only natural to wonder whether there are other options that can or should be explored. 

The purely circular orbit model is an obvious alternative: since circular orbits do not pollute the density at smaller radii, circular orbit models in principle work for all possible density profiles \citep{1984ApJ...286...27R, 2008gady.book.....B}. There are other, more general, alternatives, however.

\subsubsection{Tangential Cuddeford orbital structure}

One interesting alternative, or rather generalisation of the TOM orbital structure, is what we can refer to as the tangential Cuddeford (hereafter TC) orbital structure. This orbital structure, presented in \citetalias{2022MNRAS.512.2266B} for the special case of the uniform density sphere, is an extension of the type of dynamical models proposed by \citet{1991MNRAS.253..414C} to tangentially anisotropies. The TC orbital structure is characterised by the anisotropy profile 
\begin{equation}
\beta(r) =  \beta_0 - (1-\beta_0) \left(\frac{r^2}{\rT^2-r^2}\right).
\label{TCbeta}
\end{equation}
For each choice of the parameter $\beta_0< 1$, models with this orbital structure are indeed completely tangential at the truncation radius. The TC orbital structure is clearly a generalisation of the TOM one, which just corresponds to $\beta_0=0$. Models with $\beta_0<0$ are tangentially anisotropic in the central regions and systematically get even more tangential for increasing radius. Models with $\beta_0>0$ are radially anisotropic in the central regions, and they first become isotropic and subsequently tangentially anisotropic when moving to larger radii. 

The distribution function corresponding to the TC orbital structure has the general form
\begin{equation}
f(\calE,L) = L^{-2\beta_0}\,f_{\scriptscriptstyle{\text{C}}}(Q),
\end{equation}
with $Q$ given by Eq.~(\ref{Q-}) and $f_{\scriptscriptstyle{\text{C}}}(Q)$ a function that can be obtained from the density using an Eddington-like inversion formula (see \citetalias{2022MNRAS.512.2266B} for details). Whether or not a TC orbital structure is a viable option for a given truncated density model depends on the specific shape of the density profile, on the value of the truncation radius, and on the value of the central anisotropy. In particular, the central density slope--anisotropy inequality \citep{2006ApJ...642..752A} needs to be satisfied as a minimum requirement to have a positive distribution function. For the uniform density sphere, all TC models with $\beta_0\leqslant0$ are consistent, whereas models with radially anisotropic central regions ($\beta_0>0$) are not, as expected based on the central density slope--anisotropy inequality and as demonstrated explicitly in \citetalias{2022MNRAS.512.2266B}. A similar consistency analysis can in principle be performed for other radially truncated models. 

\subsubsection{Other orbital structures with tangential anisotropy at the truncation radius}

As far as we are aware, the TC orbital structure, with the TOM as a special case, is the only orbital structure that becomes completely tangentially anisotropic at the truncation radius and that allows for a direct calculation of the distribution function using an Eddington-like inversion formula. It is possible to construct dynamical models with other, more general anisotropy profiles that become purely tangential, but in those cases the calculation of the distribution function is far more involved. 

One possible way forward is an approach based on separable augmented densities. For any given potential--density pair, there are infinitely many different options to create augmented densities. Each choice completely determines an independent dynamical model, and there is a one-to-one correspondence between the augmented density and the distribution function \citep{1962MNRAS.123..447L, 1986PhR...133..217D, 1993MNRAS.262..401H}. A major advantage of using separable augmented densities, i.e., functions of the form
\begin{equation}
\tilde\rho(\Psi,r) = \varrho(\Psi)\,g(r),
\label{varrhog}
\end{equation}
is that the anisotropy profile of the resulting dynamical model only depends on the function $g$ in a simple way \citep[e.g.,][]{1995A&AT....7..201Q},
\begin{equation}
\beta(r) = - \frac12\,\frac{\txd \ln g}{\txd \ln r}(r).
\end{equation}
This characteristic makes it possible to set up models with a preset density profile and a preset anisotropy profile. In particular, we can aim at dynamical models that are completely tangential at $r=\rT$ and which might be suitable options for radially truncated models. Taking inspiration from \citet{2007A&A...471..419B}, we can consider the function
\begin{equation}
g(r) = \left(\frac{r}{\rT}\right)^{-2\beta_0} \left(1-\frac{r^{2\delta}}{\rT^{2\delta}}\right)^{-\xi/\delta},
\label{BVH2g}
\end{equation}
with parameters $\beta_0<1$, $\xi>0$, and $\delta>0$. The corresponding anisotropy profile for $r\leqslant\rT$ is 
\begin{equation}
\beta(r) = \beta_0 - \xi \left(\frac{r^{2\delta}}{\rT^{2\delta} - r^{2\delta}}\right),
\label{BVH2}
\end{equation}
which becomes completely tangential at the truncation radius. Comparing this to the expressions~(\ref{TOMbeta}) and (\ref{TCbeta}), we see that it reduces to the TC orbital structure for the special case $\xi=1-\beta_0$ and $\delta=1$, and to the TOM orbital structure for $\beta=0$ and $\xi=\delta=1$.

For a given density profile, the calculation of the augmented density is usually straightforward. The main challenge is the calculation of the corresponding distribution function. In general, the inversion from augmented density to distribution function involves the combination of a forward and an inverse Laplace--Mellin transform \citep{1986PhR...133..217D}. For the case of separable augmented densities, the Laplace and Mellin transforms can be separated, but the entire inversion still remains a significant challenge. One characteristic we can exploit is that the inversion is a linear operation. If we manage to expand the factor $\varrho(\Psi)$ in the augmented density~(\ref{varrhog}) as a linear combination of simpler components,
\begin{equation}
\varrho(\Psi) = \sum_k c_k \varrho_k(\Psi),
\end{equation}
and for each component $\tilde\rho_k(\Psi,r) = \varrho_k(\Psi)\,g(r)$ we can calculate the corresponding distribution function $F_k(\calE,L)$, then we immediately have the distribution function for the complete model, 
\begin{equation}
f(\calE,L) = \sum_k c_k F_k(\calE,L).
\end{equation}
This approach has been applied by \citet{2009ApJ...690.1280V} to construct distribution functions with a general anisotropy profile for the set of realistic dark matter halo models proposed by \citet{2005MNRAS.363.1057D}, based on the library of analytical components set up by \citet{2007A&A...471..419B}. In principle, this approach can also be followed for the radially truncated density models considered in this paper if one wants to generate models with a more general anisotropy profile than the TOM or TC ones.

\subsubsection{Purely radial orbital structure}

Up to now we have focused on orbital structures that become completely tangential at the truncation radius as candidates to support truncated density models. There is, however, another possibility: orbital structures that are completely radial at the truncation radius. The simplest example of an orbital structure with this characteristic is the one in which only purely radial orbits are populated. For a purely radial orbital structure, the distribution function has the form
\begin{equation}
f(\calE,L) = \delta(L^2)\,f_{\text{rad}}(\calE),
\end{equation}
in which $f_{\text{rad}}(\calE)$ can be obtained from the density using an Eddington-like inversion formula that only involves a single differential \citep[e.g.,][]{1984ApJ...286...27R, 2016MNRAS.462..298O, 2018MNRAS.473.5476C}. In fact, the purely radial orbital structure can be considered as a special limiting case of the TC orbital structure for $\beta_0\to1$.

The idea that a purely radial orbital structure can support a truncated density model is illustrated by the example of the truncated singular isothermal sphere, which has the density profile
\begin{equation}
\rho(r) = \frac{M}{4\pi\,\rT\,r^2}\,\Theta(\rT-r).
\end{equation}
This simple model can be supported self-consistently by a purely radial orbital structure, as demonstrated by \citet{1984sv...bookQ....F}. Interestingly, the truncated singular isothermal sphere can be supported by the purely radial ($\beta=1$) and the purely circular ($\beta=-\infty)$ orbital structure, but not by any constant anisotropy model with $-\infty<\beta<1$, including the ergodic model ($\beta=0$).

\subsection{Connection to real dynamical systems}

The study presented here is primarily a theoretical study on the existence and consistency of radially truncated dynamical models. In this subsection we discuss two aspects of our models in relation to real dynamical systems.

\subsubsection{The TOM dynamical structure}

We have demonstrated in this paper that many spherical density profiles with a discontinuous radial truncation can be supported self-consistently by a TOM orbital structure. It remains to be seen whether such a TOM orbital structure can be representative for the orbital structure of real dynamical systems. 

Observed galaxy clusters and simulated dark matter haloes tend to have an anisotropy profile that is roughly isotropic at the central and mildly to significantly radial at larger radii \citep{2001ApJ...563..483T, 2004MNRAS.352..535D, 2011MNRAS.415.3895L, 2012ApJ...752..141L, 2013MNRAS.434.1576W, 2016MNRAS.462..663B, 2021MNRAS.500.3151S}. The anisotropy profiles of observed galaxies show a larger variety. A detailed study of the internal kinematics of 161 passive galaxies by \citet{2022ApJ...930..153S} showed a variety of anisotropies, with flatter galaxies on average more tangential and more massive and round galaxies more radially anisotropic. Estimates of the anisotropy of galaxies at large radii based on the orbits of globular clusters also vary widely, from mildly radially anisotropic for the Milky Way \citep{2018A&A...616A..12G} and NGC\,5846 \citep{2014MNRAS.439..659N} to roughly isotropic for M87 \citep{2014ApJ...792...59Z} to significantly tangential for NGC\,1407 \citep{2012MNRAS.423.2177S}. It needs to be added that these different estimates are often subject to significant systematic uncertainties. 

The radially truncated models and the TOM orbital structure discussed in this paper might be most relevant in the context of globular clusters. Using detailed N-body simulations, \citet{2003MNRAS.340..227B} and \citet{2015MNRAS.451.2185S} argue that globular clusters are nearly isotropic in the inner regions, but that their outer parts rapidly become tangentially anisotropic. \citet{2014MNRAS.443L..79V} found a similar orbital structure for simulated isolated globular clusters: an inner isotropic core, followed by a region of increasing radial anisotropy. For clusters evolving in an external tidal field, however, they found that the tangential anisotropy peaks in the cluster intermediate regions and then progressively decreases, with the cluster outermost regions being characterised by isotropy or a mild tangential anisotropy. \citet{2017MNRAS.471.1181B} found an even larger diversity in the orbital structure of their simulated globular clusters: at small radii, all globular clusters tend to be isotropic, but they can be either radially or tangentially anisotropy in the intermediate and outer regions, with the velocity anisotropy primarily depending on the strength of the tidal field cumulatively experienced by a cluster. 

Detailed observations and dynamical models for nearby globular clusters also show a range of orbital structures. The best-fit Schwarzschild orbit-superposition model for $\omega$~Cen by \citet{2006A&A...445..513V} is close to isotropic in the inner regions and becomes increasingly tangentially anisotropic in the outer regions. In a study of nine inner Milky Way globular clusters, \citet{2021AJ....161...41C} found that most of their clusters are isotropic even out to their half-light radii, while NGC\,6380 was found to be tangentially anisotropic beyond its half-light radius. 

\subsubsection{Sharp versus smooth truncation}

The goal of this series of papers is to study the existence of self-consistent dynamical models with a preset density distribution with a finite extent. In the present paper we focused on such models created by truncating smooth infinite-extent models abruptly at a truncation radius. This results in models with a sharp and discontinuous density truncation, with $\rho(r) \ne 0$ as $r\to\rT$. Such a discontinuous density truncation is the simplest mathematical construction to generate a truncation, but it is probably not so realistic when considering real dynamical systems. Instead of a sharp and discontinuous truncation one could consider a more gradual truncation; this can be realised by replacing the Heaviside step function in Eq.~(\ref{truncdensity}) by a continuous function that smoothly tends to zero for $r\to\rT$ and leaves the density profile unaffected for $r\ll\rT$. This makes the mathematics of the problem significantly more complex, but one might wonder how it would affect the conclusions of this work.

One of the immediate consequences of a discontinuous density truncation is that such models can never be supported by an ergodic orbital structure, as demonstrated mathematically in Section~{\ref{iso.sec}} and discussed in Section~{\ref{interpret.sec}}. Based on expression~(\ref{dftrunc}) one would be inclined to argue that, for $\rhoT=0$, the negativity of the ergodic distribution function for $\calE\to\calET$ is no longer a certainty, and thus that smoothly truncated models with an ergodic orbital structure might be consistent. 

This situation reminds of the consistency analysis of the family of broken power-law models, in which the density is a combination of two pure power laws merged the break radius \citep{2020ApJ...892...62D}. \citet{2021MNRAS.503.2955B} demonstrated that broken power-law models can never be supported by an ergodic orbital structure, because the sharp break in the density profile at the break radius forces the ergodic distribution function to be negative at binding energies just beyond $\calE_{\text{b}} = \Psi(\rb)$. Instead of the infinitely sharp density break that is a defining property of the double power-law models, one can consider a more gradual transition between the two power laws. This yields the family of double power-law models \citep{1996MNRAS.278..488Z}, where an additional parameter controls the smoothness of the transition between the inner and outer power-law density profile. For sufficiently soft transitions, these double power-law models can be supported by an ergodic orbital distribution. For sufficiently sharp transitions, however, we encounter the same situation as for the broken power-law models: the distribution function is negative for $\calE\gtrsim\calE_{\text{b}}$ and the ergodic model is inconsistent.

This case suggests that the replacement of the infinitely sharp truncation by a more gentle one, with a truncated but continuous density profile as a result, could result in a situation where consistent ergodic orbital structures are possible as long as the truncation is sufficiently soft. For truncations that are sufficiently sharp, one may expect the ergodic distribution function, and constant-anisotropy distribution functions, to remain inconsistent. An in-depth investigation is beyond the scope of this paper.
 
\section{Summary}
\label{summary.sec}

This paper is the second in a series on the search for and the discussion of self-consistent dynamical models with a finite extent. In the first paper of this series \citepalias{2022MNRAS.512.2266B}, we performed an in-depth study of the uniform density sphere, the simplest model with a radially truncated density profile. By explicitly calculating the phase-space distribution function, we demonstrated that the uniform density sphere cannot be supported by an ergodic, constant anisotropy, or radial Osipkov--Merritt orbital structure. On the other hand, we showed that it can be supported by a TOM orbital structure, and more generally, by a TC orbital structure as long as the central anisotropy is not radial. In this second paper, our ambition was to investigate in a more systematic way by which orbital structures radially truncated density models can be supported.

The first conclusion of this paper is that no radially truncated spherical model with a density discontinuity at the truncation radius can be supported by an ergodic orbital structure. This can be shown explicitly in a relatively straightforward way by calculating the ergodic distribution function. The same conclusion accounts for all constant anisotropy or Osipkov--Merritt orbital structures: none of these orbital structures can support a radially truncated model with a density discontinuity.

Our second conclusion is that the TOM orbital structure with $\ra=\rT$ is capable of self-consistently supporting a large set of truncated density models. We formulate a consistency hypothesis: if a spherical density model can be supported self-consistently by an ergodic orbital structure, then the model that is radially truncated at $r = \rT$ can be supported self-consistently by a TOM orbital structure with $\ra = \rT$. The validity of this consistency hypothesis is not formally proven, but it is corroborated by an extensive suite of numerical tests.

These results can be understood by considering a dynamical models as a superposition of orbits. The combination of a sharp density truncation and the requirement of velocity isotropy at the break radius can only be realised by an infinite number of orbits with binding energy $\calET$, which generates a mass density excess at smaller radii that can only be compensated by orbits with negative weights. The result is a negative, and thus non-physical ergodic phase-space distribution function. 

These conclusions demonstrate that the uniform density is not a special case, but rather a typical example of a truncated density model. We hope this paper is another step towards a broader understanding of the dynamical structure of models with a finite extent.

\section*{Acknowledgements}

MB warmly thanks Luca Ciotti and Herwig Dejonghe for discussions that led to an improved version of this work. The referee, Carlo Nipoti, is acknowledged for his insightful comments and suggestions. For this work we made use of the Python packages {\tt{matplotlib}} \citep{2005ASPC..347...91B} and {\tt{SciPy}} \citep{2020NatMe..17..261V}.

\section*{Data availability}

No astronomical data were used in this research. The data generated and the plotting routines will be shared on reasonable request to the corresponding author. The {\tt{SpheCow}} code \citep{2021A&A...652A..36B} is publicly available on GitHub.\footnote{\url{https://github.com/mbaes/SpheCow}}

\bibliographystyle{mnras}
\bibliography{Truncation_bib}

\appendix
\section{TOM orbital structure in {\tt{S{\lowercase{phe}}C{\lowercase{ow}}}}}
\label{TOM.sec}

For the numerical calculation of the dynamical structure of an arbitrary spherical model under the assumption of a TOM orbital distribution, we needed to implement the relevant formulae in {\tt{SpheCow}}. The TOM orbital structure is characterised by an anisotropy parameter $\ra$, and it is only physically meaningful for spherical density profiles with a finite extent and with $\ra\geqslant\rT$. In the remainder of this section, it is assumed that the density profile $\rho(r)$ satisfies this condition. It is also assumed that the first and second derivatives of the density have been implemented, or that they can be computed from the surface density and its derivatives. 

The radial velocity dispersion is most easily obtained as the solution of the Jeans equation, which becomes for the TOM orbital structure \citep{1985AJ.....90.1027M, 2005MNRAS.362...95M},
\begin{equation}
\sigma_r^2(r) = \frac{1}{\varrho(r)} \int_r^{\ra} \frac{\varrho(u)\,GM(u)\,\txd u}{u^2},
\label{sigmarTOM}
\end{equation}
with 
\begin{equation}
\varrho(r) = \left(1-\frac{r^2}{\ra^2}\right) \rho(r).
\end{equation}
The tangential velocity dispersion is 
\begin{equation}
\sigma_\txt^2(r) = 2\left[1-\beta(r)\right] \sigma_r^2(r),
\end{equation}
with the anisotropy profile $\beta(r)$ given by expression~(\ref{TOMbetagen}). The general expression for the projected velocity dispersion for an anistropic spherical model is \citep{1982MNRAS.200..361B, 1987MNRAS.224...13D}
\begin{equation}
\Sigma(R)\,\sigma_\txp^2(R) = 2 \int_R^\infty \left[ 1 - \frac{R^2}{u^2}\,\beta(u)\right] \frac{\rho(u)\,\sigma_r^2(u)\,u\,\txd u}{\sqrt{u^2-R^2}},
\end{equation}
which becomes for the TOM orbital structure,
\begin{equation}
\Sigma(R)\,\sigma_{\text{p}}^2(R) 
= 
2 \int_R^{\ra}
\left(\frac{\ra^2-u^2+R^2}{\ra^2-u^2}\right) 
\frac{\rho(u)\,\sigma_r^2(u)\,u\,\txd u}{\sqrt{u^2-R^2}}.
\end{equation}
An equivalent expression, which eliminates one level of quadrature, can be obtained by substituting expression~(\ref{sigmarTOM}) in this expression and changing the order of integration \citep{2005MNRAS.363..705M, 2014MNRAS.442.3284A, 2021isd..book.....C}. The result is
\label{sigmap_om}
\begin{equation}
\Sigma(R)\,\sigma_{\text{p}}^2(R) 
= 
\int_R^{\ra} \frac{w(u,R)\,\rho(u)\,GM(u)\,\txd u}{u^2},
\end{equation}
with
\begin{multline}
w(u,R) = \left(\frac{\ra^2-u^2}{\ra^2-R^2}\right)
\\
\times
\left( 
\frac{2\ra^2-R^2}{\sqrt{\ra^2-R^2}}
\arctanh\!\sqrt{\frac{u^2-R^2}{\ra^2-R^2}}
+ \frac{R^2\sqrt{u^2-R^2}}{\ra^2-u^2} 
\right).
\end{multline}
The TOM distribution function is obtained using Eq.~(\ref{om}). We recast this expression to a form that is more suitable for numerical integration,
\begin{equation}
f(Q) = \frac{1}{2\sqrt2\,\pi^2} 
\left[
\frac{2\ra\,\rho(\ra)}{GM_{\text{tot}}\, \sqrt{Q-\Psi(\ra)}}
+
\int_{r(Q)}^{\ra} \frac{\Delta(u)\,\txd u}{\sqrt{Q-\Psi(u)}}
\right],
\label{fQDeltaQ}
\end{equation}
with 
\begin{equation}
\Delta(r) = \frac{r^2}{GM(r)}\left[\varrho''(r) + \varrho'(r)
\left(\frac{2}{r} - \frac{4\pi\,\rho(r)\,r^2}{M(r)}\right)
\right].
\label{defDeltaQ}
\end{equation}
Finally, for the pseudo-differential energy distribution \citep{1991MNRAS.253..414C, 2021A&A...652A..36B}, i.e., the distribution of mass as a function of $Q$, we have
\begin{equation}
\calN(Q) = f(Q)\,g(Q),
\end{equation}
with $g(Q)$ a pseudo-density-of-states function, 
\begin{equation}
g(Q) 
=
16\sqrt2\,\pi^2 \int_0^{r(Q)} \left(1-\frac{u^2}{\ra^2}\right)^{-1} \sqrt{\Psi(u)-Q}\,u^2\,\txd u.
\end{equation}

\bsp	
\label{lastpage}
\end{document}